\documentclass[aps,prb,amsmath,twocolumn,amssymb,floatfixng,showpacs,
superscriptaddress,footinbib]{revtex4-1}
\pdfoutput=1
\usepackage[dvips]{graphics}
\usepackage{bm}
\usepackage{float}
\usepackage{epsfig}
\usepackage{enumerate}
\usepackage{color}
\usepackage{graphicx}
\usepackage[colorlinks=true,linktoc=page,linkcolor=magenta,citecolor=cyan]{hyperref}

\newcommand{\etal}{{\it et al.~}}
\newcommand{\ie}{{\it i.e., }}
\newcommand{\eg}{{\it e.g. }}
\newcommand{\ua}{{\uparrow }}
\newcommand{\da}{{\downarrow }}

\newcommand\bea{\begin{eqnarray}}
\newcommand\eea{\end{eqnarray}}
\newcommand\beq{\begin{equation}}  
\newcommand\eeq{\end{equation}}

\newcommand{\non}{\nonumber}  
\begin{document} 
\title{Spin selective coupling to Majorana zero modes in mixed singlet and triplet superconducting nanowire} 

\author{Ganesh C. Paul}
\email{ganeshpaul@iopb.res.in}
\affiliation{Institute of Physics, Sachivalaya Marg, Bhubaneswar-751005, India}
\affiliation{Homi Bhabha National Institute, Training School Complex, Anushakti Nagar, Mumbai 400085, India}
\author{Arijit Saha}
\email{arijit@iopb.res.in}
\affiliation{Institute of Physics, Sachivalaya Marg, Bhubaneswar-751005, India}
\affiliation{Homi Bhabha National Institute, Training School Complex, Anushakti Nagar, Mumbai 400085, India}
\author{Sourin Das}
\email{sdasdu@gmail.com}
\affiliation{Department of Physical Sciences, IISER Kolkata, Mohanpur, West Bengal 741246, India}
\affiliation{Department of Physics and Astrophysics, University of Delhi, Delhi 110007, India}

\begin{abstract}
We theoretically investigate the transport properties of a quasi one dimensional ferromagnet-superconductor junction where the superconductor consists of mixed singlet and triplet pairings. We show that
 the relative orientation of the stoner field ($\bf{\tilde{h}}$) in the ferromagnetic lead and the $\bf{d}$ vector of the superconductor acts like a on-off switch for the zero bias conductance of the device.
 In the regime, where triplet pairing amplitude dominates over the singlet counterpart (topological phase), a pair of Majorana zero modes appear at each end of the superconducting part of the nanowire.
When $\bf{\tilde{h}}$ is  parallel or anti-parallel to the $\bf{d}$ vector, transport gets completely blocked due to blockage in pairing while, when $\bf{\tilde{h}}$ and $\bf{d}$ are perpendicular to 
each other, the zero energy two terminal differential conductance spectra exhibits sharp transition from $4e^2/h$ to $2e^2/h$ as the magnetization strength in the lead becomes larger than the 
chemical potential indicating the spin selective coupling of pair of Majorana zero modes to the lead. 
\end{abstract}

\maketitle
%----------------------------------------------------------------------
\section{Introduction}{\label{sec:I}}
%----------------------------------------------------------------------
 Localized Majorana zero modes (MZMs) that appear at the end of one dimensional topological superconductor are anticipated to be the building blocks of future topological quantum computers~\cite{kitaev2001unpaired,nayak2008non,alicea2012new,beenakker2013search,elliott2015colloquium,ramonaquado2017}. Theoretical proposals ~\cite{oreg2010helical,lutchyn2010majorana} to engineer such a topological superconductor from a semiconducting nanowire (NW) with Rashba spin-orbit coupling have stimulated a lot of recent exciting experiments towards realizing this exotic phase hosting Majorana zero mode (MZM). The zero bias peak (ZBP) in the differential conductance was predicted~\cite{bolech2007observing,law2009majorana,flensberg2010tunneling,wimmer2011quantum} in hybrid superconductor-semiconductor systems. However, the earler experimental findings~\cite{mourik2012signatures,das2012zero,rokhinson2012fractional,finck2013anomalous} were largely debated because of the possibility of ZBP appearing from coalescing Andreev levels~\cite{kells2012near}, Kondo physics~\cite{lee2012zero,finck2013anomalous}, weak antilocalization~\cite{pikulin2012zero}, disorder~\cite{bagrets2012class} or multi band effects~\cite{liu2012zero} etc. With the improved devices, more recent experiments~\cite{albrecht2016exponential,deng2016majorana,jeon2017distinguishing} reveal more convincing signatures of 
MZMs. Nevertheless, the topological origin of ZBP is not ensured as the ZBP appearing from Andreev bound levels also mimic those of MZMs~\cite{liu2017andreev,ptok} and it is hard to distinguish them. 
Hence, newer probes of MZMs beyond ZBP is of great importance~\cite{prada2017measuring,cayao2017majorana,hell2017distinguishing,aptok,hoffman2017spin}.

In this article we show that it is possible to use a ferromagnetic lead for probing MZM hosted in non-centrosymmtric superconductor (NCS)~\cite{bauer2012non,yip2014noncentrosymmetric} which has broken inversion symmetry and the corresponding pair potential is composed of both singlet and triplet spin states ~\cite{sigrist1991phenomenological}. Earlier zero energy peak 
in transport accross normal-metal-NCS junction was reported~\cite{honerkamp1998andreev,tanaka2010anomalous,burset2014transport,ytanaka} without establishing it's connection to the MZM. Here we investigate the properties of two-terminal conductance of a quasi one dimensional ferromagnet-superconductor (FS) junction where the superconductor lacks inversion symmetry. We employ extended Blonder-Tinkham-Klapwijk (BTK) formalism~\cite{blonder1982transition} for our analysis. We show that in the topological phase, when triplet pairing dominates over the singlet one, the two terminal differential conductance at zero bias exhibits a sharp transition from $4e^2/h$ to $2e^2/h$ as the magnetization strength becomes larger than the chemical potential in the lead. We also observe a spin selective coupling of the ferromagnetic lead to the pair of MZMs as a function of the magnetization of the ferromagnet, with respect to the $\bf{d}$ vector in the topological regime.

The rest of the paper is structured as follows. In Sec.~\ref{sec:II}, we describe the model Hamiltonian for our setup and the method for obtaining the conductance. 
Sec.~\ref{sec:III} is devoted to our numerical results for the differential conductance. Finally, we summarize our results and conclude in Sec.~\ref{sec:IV}.

%----------------------------------------------------------------------------------------------------------------------------
%---------------------------------------------------------------------------------------------------------------------------- 
\begin{figure}[!thpb]
\centering
\includegraphics[width=0.9 \linewidth]{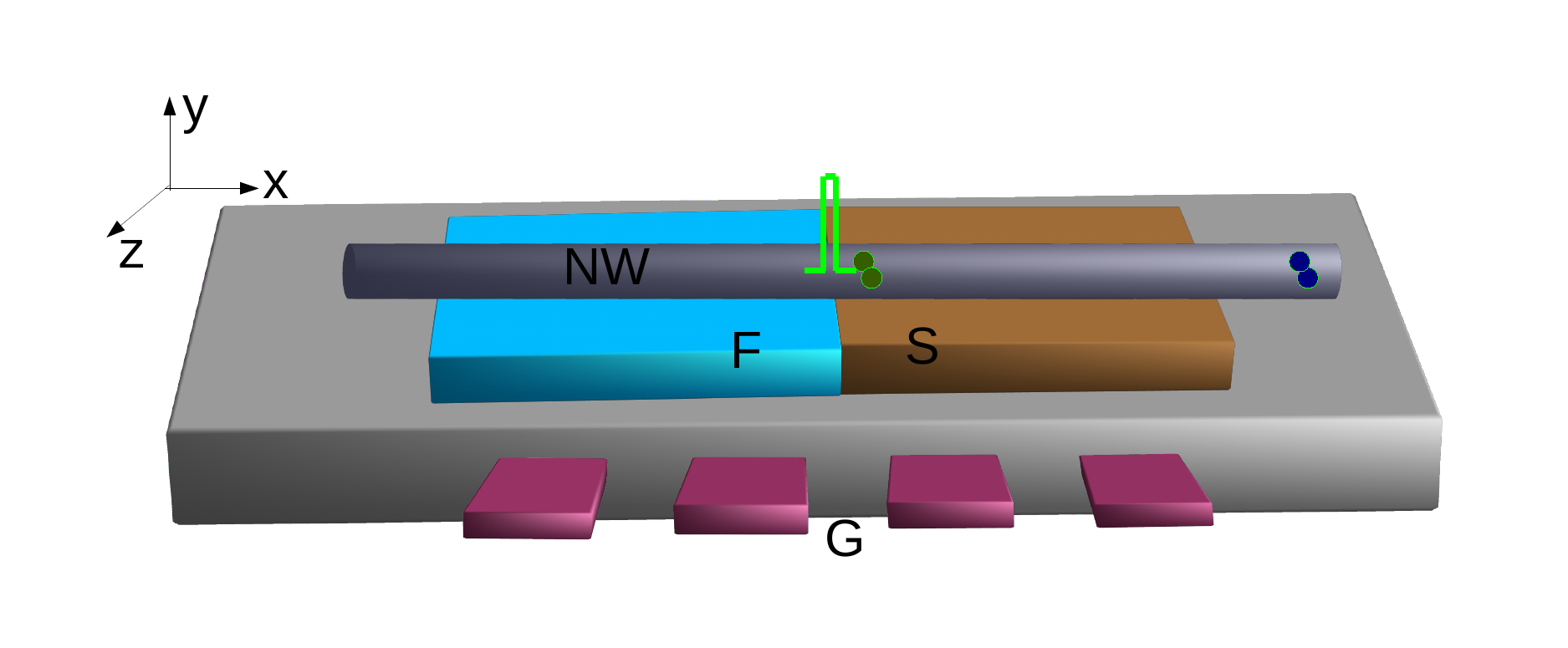}
\caption{(Color online) Schematic of our FS setup in which a quasi 1D NW (dark gray) is placed in close proximity to a ferromagnet F (light blue, light gray) 
and a bulk inversion symmetry broken superconductor S (brown, gray). Superconductivity is induced in the NW via the proximity effect. The gates $G$ (maroon, light gray) 
control the chemical potential in different regions of the NW. A $\delta$-function barrier is symbolically depicted by the light green (light gray) rectangular barrier at the FS interface. 
Two pair of MZMs are shown by green (light gray) and dark blue (gray) circles at each end of the superconducting part of the NW.}
\label{model}
\end{figure}
%-----------------------------------------------------------------------------------------------------------------------------
%-----------------------------------------------------------------------------------------------------------------------------

%----------------------------------------------------------------------
\section{Model and Method}{\label{sec:II}}
%----------------------------------------------------------------------
In Fig.~\ref{model}, we present the schematic of our proposed FS set-up in which a part of the quasi one-dimensional (1D) NW is placed in close proximity to a ferromagnet and rest to a bulk superconducting material with broken inversion symmetry. Here ferromagnetism and superconductivity are induced in the NW via the proximity effect. It is assumed that both the strength and the direction of the induced magnetization vector in the NW can be controlled via the bulk ferromagnet~\cite{hofstetter2010ferromagnetic}. The engineered FS structure is attached to a normal metal lead (not shown)
via which the current can be measured. The gate voltages (denoted by $G$) can tune the chemical potential in differnt parts of the NW. 

We choose the $x$-axis along the axis of the NW. The interface of F and S regions of the NW is taken at $x=0$ for simplicity. We consider an insulating barrier at the FS interface 
which is modeled by a $\delta$-function potential given as $V(x)=(\hbar^2k_F/m)Z\delta(x)$ where $k_F$ is the Fermi wave vector in the lead, $m$ denotes electron mass and $Z$ is the dimensionless barrier strength. Chemical potential in F and S regions are $\mu$ and $\mu+U$ respectively where $U$ is extra gate potential in the S region to tune the Fermi energy mismatch. 

In the superconducting region, which is composed of both singlet and triplet pairing states, the pairing potential 
$\hat{\Delta}(\mathbf{k})$ ($2 \times 2$ matrix), in general, can be written as $\hat{\Delta}(\mathbf{k})=i[\Delta_s(\mathbf{k})\hat{\sigma}_0 + 
\sum_{j=1}^{3}d_j(\mathbf{k})\hat{\sigma}_j]\hat{\sigma}_2 e^{i\phi}$~\cite{sigrist1991phenomenological}. Here, $\hat{\sigma}_{1,2,3}$ are Pauli spin matrices operating on spin space and $\phi$ is the superconducting phase. Throughout our analysis, we consider only the mean-field value of $\Delta_s(\mathbf{k})$ \ie $\Delta_s(\mathbf{k})=\Delta_{s}$.
In contrast, the triplet pairing potential is characterized by an odd vector function as $\mathbf{d}(\mathbf{k})=-\mathbf{d}(-\mathbf{k})$. Following Burset \etal~\cite{burset2014transport}, we consider
the chiral triplet state of the form,
$\mathbf{d}(\mathbf{k})\,=\,\Delta_p\frac{k_x+i{\chi}k_y}{\lvert \mathbf{k}\rvert}
\hat{z}\,=\,\Delta_{p}e^{i{\chi}\theta}\hat{z} \, $,
where $\Delta_{p}$ is the non-negative amplitude of the triplet pairing potential and $\chi=\pm$ denotes opposite chiralities. Here, $\chi$ determines the orientation of the angular momentum of the Cooper pairs and $\theta$ represents the relative phase between the singlet and triplet pairing states. The superconducting  pairing preserves time reversal symmetry (TRS) either for $\theta=n\pi$ or for $\theta = n\pi/2$, with $n = 0,1,. . .$\,. For 1D case, depending on the value $\theta$, the Hamiltonian can be categorized to either in class C, class D, or class DIII, according to the Altland-Zirnbauer symmetry classification~\cite {altland1997nonstandard,schnyder2008classification}. For the case with $\theta = 0$, the Hamiltonian belongs to the nontrivial DIII symmetry class if $\Delta_p>\Delta_s$~\cite{budich2013topological}. With this simplification, paring potential now takes the form, $\hat{\Delta}(\mathbf{k})=i[\Delta_s\hat{\sigma}_0+\Delta_{p}e^{i{\chi}\theta}\hat{\sigma}_3]\hat{\sigma}_{2}e^{i\phi}$. 

The FS junction can be described by the Bogoliubov-deGennes (BdG) equations as,
%\bea
$H(\mathbf{k}) \Psi(\mathbf{k}) = \epsilon \Psi(\mathbf{k})$
%\eea
where the Hamiltonian $H(\mathbf{k})$ can be written as
\footnotesize
\setlength{\arraycolsep}{2.5pt} % default: 5pt
\medmuskip = 1mu % default: 4mu plus 2mu minus 4mu
\begin{align}
\begingroup
\begin{pmatrix}
\small
\centering
E(\mathbf{k})-\tilde{h}\cos{\psi} & -\tilde{h}\sin{\psi}\,e^{-i\phi_F} & 0 & \Delta_{+}\\
-\tilde{h}\sin{\psi}\,e^{i\phi_F} & E(\mathbf{k})+\tilde{h}\cos{\psi} & -\Delta_{-} & 0\\
0 & \Delta_{-}^{*} & -E(\mathbf{-k})+\tilde{h}\cos{\psi} & +\tilde{h}\sin{\psi}\,e^{-i\phi_F}\\
\Delta_{+}^{*} & 0 & \tilde{h}\sin{\psi}\,e^{i\phi_F} & -E(\mathbf{-k})-\tilde{h}\cos{\psi}\nonumber
\end{pmatrix}
\endgroup
\end{align}

\normalsize
with $E(\mathbf{k})=k^2/2-\mu$ and $\Delta_{\pm}=[\Delta_s \pm \Delta_{p}e^{i{\chi}\theta}]e^{i\phi}$ . 
%--------------------------------------------------------------------------------------------------
%---------------------------------------------------------------------------------------------------
\begin{figure}[!thpb]
\centering
\includegraphics[width=0.85 \linewidth]{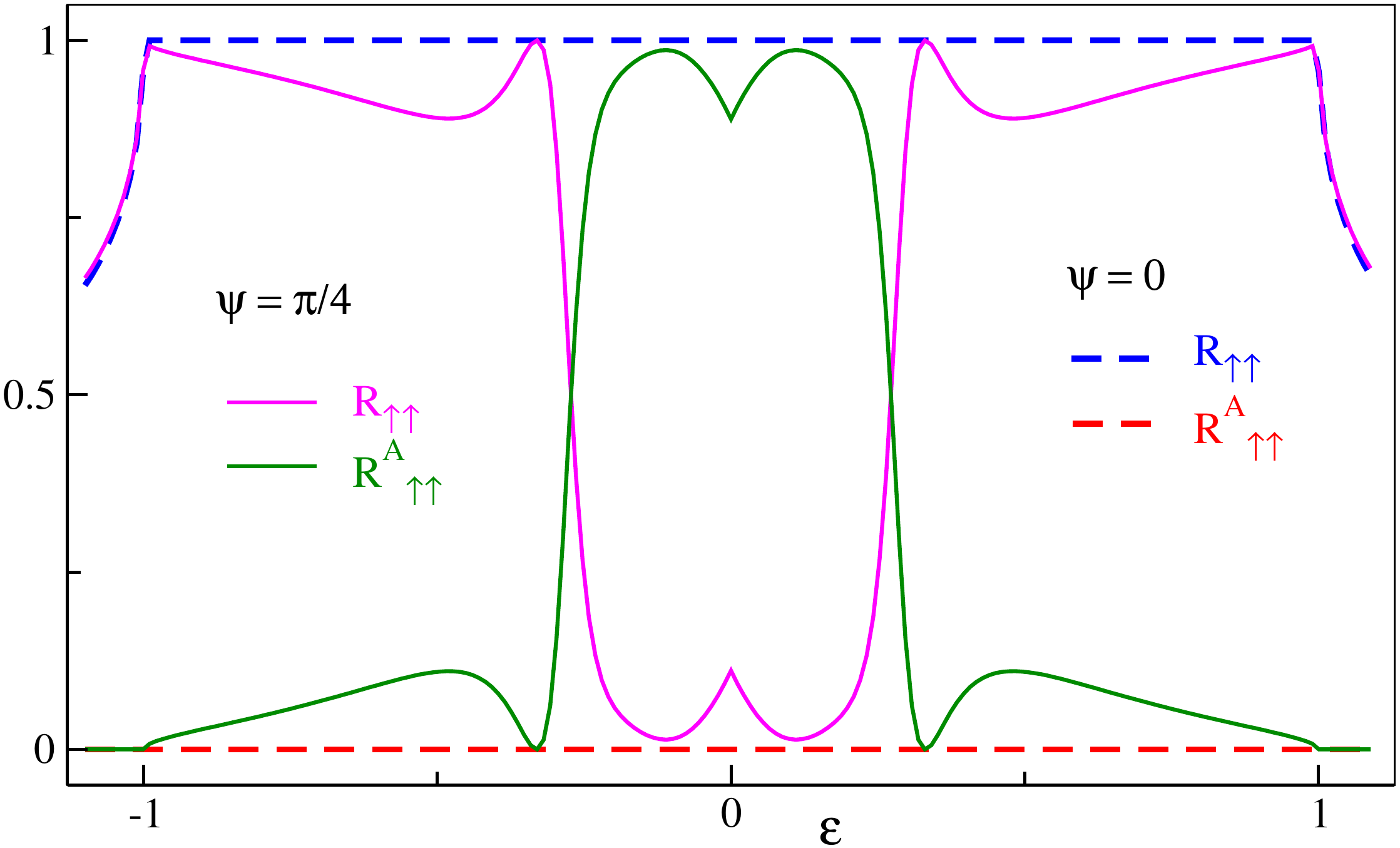}
\caption{(Color online) The behavior of reflection and SFAR scattering probabilities are shown as a function of incident electron energy (normalized by $(\Delta_p+\Delta_s)$) for two different angle of magnetization $\psi$ of the ferromagnet. The value of the other parameters are chosen as $\tilde{h}=1$, $\mu=0$, $\Delta_s=0$, $\Delta_p=1$, $Z=4$ and $U=15$.}
\label{GvsE}
\end{figure}
%--------------------------------------------------------------------------------------------------
%--------------------------------------------------------------------------------------------------

We consider the band energy $E({\pm \mathbf{k}})$ of the NW as $E(\pm k_x)$ for a particular choice of $k_y$ in the quasi 1D limit. We assume a situation where the transverse confining potential and the chemical potential in the NW are tuned such that only the lowest sub-band is participating ($k_{y}=0$ mode) in transport and hence $\theta=0$. $\theta\neq0$ corresponds to different symmetry class.
It is assumed that the band energies for the electrons moving to the left and right are equal to each other. We define right movers by $\theta_{+}=\theta$ and left movers by $\theta_{-}=\pi-\theta$ which reduces to $\theta_{+}=0$ and $\theta_{-}=\pi$ in our case. The effective pairing potential depends on different spin channels as well as the direction of motion. Right movers with spin $\ua$ and $\da$ are effected by the pairing potential $\Delta_{+}(\theta_{+})$ and $-\Delta_{-}(\theta_{+})$ respectively while left movers sense $\Delta_{+}(\theta_{-})$ and $-\Delta_{-}(\theta_{-})$ corresponding to $\ua$ and $\da$ spin channels respectively~\cite{burset2014transport,paul2016transport}. 

The magnetization vector in the ferromagnetic region is considered to be $\bold{\tilde{h}}= \tilde{h}\{\sin{\psi}\cos{\phi_F},\sin{\psi}\sin{\phi_F},\cos{\psi}\} $. Here, $\tilde{h}$ is the strength of the 
magnetization vector and $\psi$, $\phi_F$ are the polar and azimuthal orientation angle respectively. In the F region ($x<0$), $\Delta_{\pm}=0$. On the other hand, in the superconducting side ($x>0$), 
$\tilde{h}=0$. 
Wave functions inside the F region are given by
\bea
\Psi_{F\ua}^e=\{e^{-i\phi_F}\cos{\psi/2},\sin{\psi/2},0,0\}^T, \non \\
\Psi_{F\da}^e=\{-e^{-i\phi_F}\sin{\psi/2},\cos{\psi/2},0,0\}^T,\non \\
\Psi_{F\ua}^h=\{0,0,e^{i\phi_F}\cos{\psi/2},\sin{\psi/2}\}^T, \non \\
\Psi_{F\da}^h=\{0,0,-e^{i\phi_F}\sin{\psi/2},\cos{\psi/2}\}^T\ .
\eea
For an incoming electron with spin $\sigma$, total wavefunction in F region becomes,
$\Psi_F=\Psi_{F\sigma}^e\,e^{ik_{F\sigma}^{e}x}+r_{\sigma\sigma}\Psi_{F\sigma}^e\,e^{-ik_{F\sigma}^{e}x}
+r_{\sigma,-\sigma}\Psi_{F-\sigma}^e\,e^{-ik_{F-\sigma}^{e}x}+r_{\sigma\sigma}^A\Psi_{F\sigma}^h\,e^{ik_{F\sigma}^{h}x}
+r_{\sigma-\sigma}^A\Psi_{F-\sigma}^h\,e^{ik_{F-\sigma}^{h}x}.
$
Here $r_{\sigma-\sigma}^A$ and $r_{\sigma\sigma}^A$ denote the amplitudes for the conventional Andreev reflection (AR) and spin-flip Andreev reflection (SFAR)~\cite{hogl} while $r_{\sigma\sigma}$ 
and $r_{\sigma-\sigma}$ correspond to the normal and spin flip reflection amplitudes respectively. 

%------------------------------------------------------------------------------------------
%------------------------------------------------------------------------------------------
\begin{figure}[!thpb]
\centering
\includegraphics[width=1.0 \linewidth]{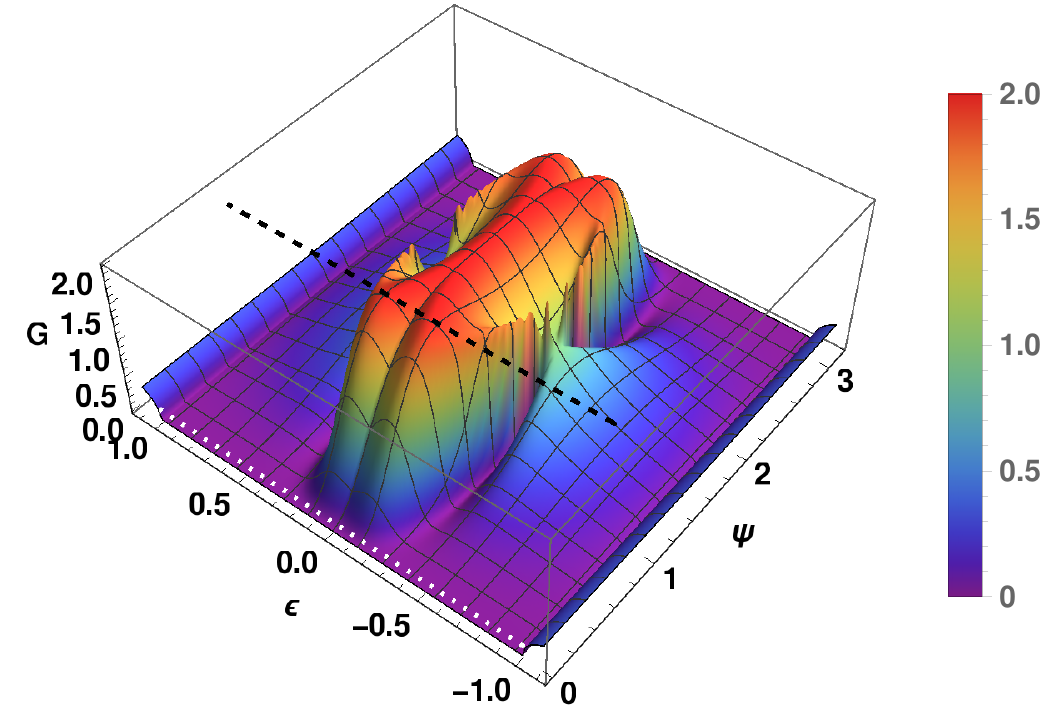}
\caption{(Color online) The behavior of differential conductance G (in unit of $e^2/h$) is displayed in the plane of incident electron energy ($\epsilon$) and angle of magnetization ($\psi$). 
The values of the other parameters are chosen to be the same as mentioned in Fig.~\ref{GvsE}. Here, black dashed and white dotted lines correspond to the behavior of the same for $\psi=\pi/4$ 
and $\psi=0$ respectively \ie they highlight $G$ for Fig.~\ref{GvsE}.
}
\label{GvsEpsi}
\end{figure}
%----------------------------------------------------------------------------------------
%----------------------------------------------------------------------------------------

%-----------------------------------------------------------------------------------------------------------------------
%-----------------------------------------------------------------------------------------------------------------------
\begin{figure*}[!thpb]
\hspace*{\fill}%
\includegraphics[width=0.5 \linewidth]{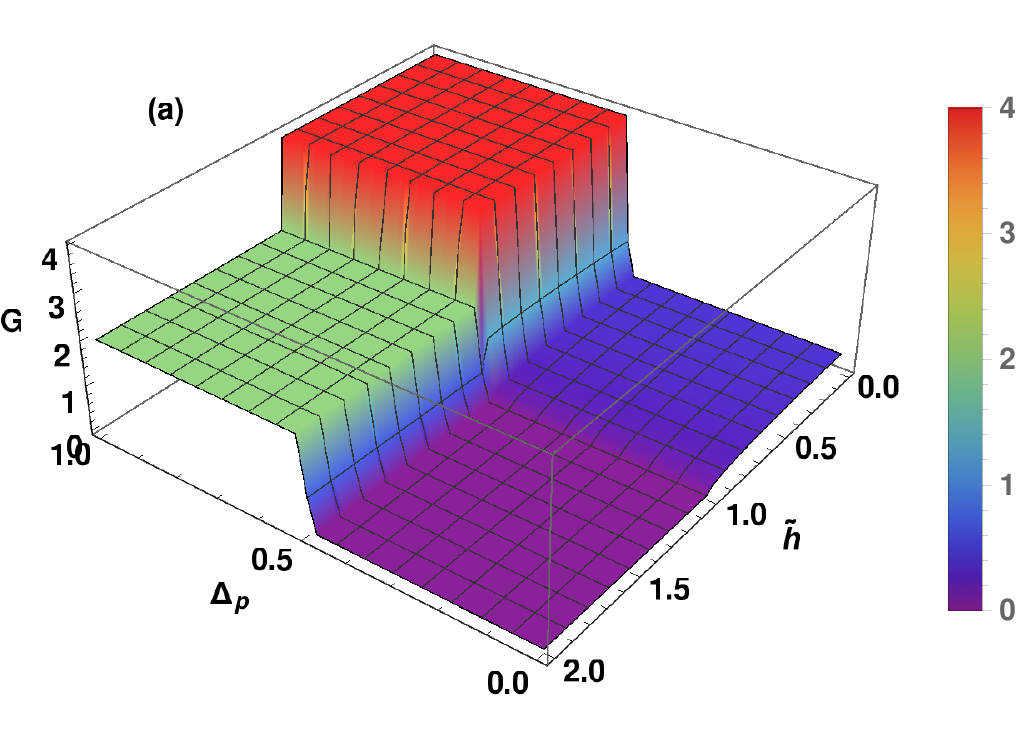}\hfill%
\includegraphics[width=0.5 \linewidth]{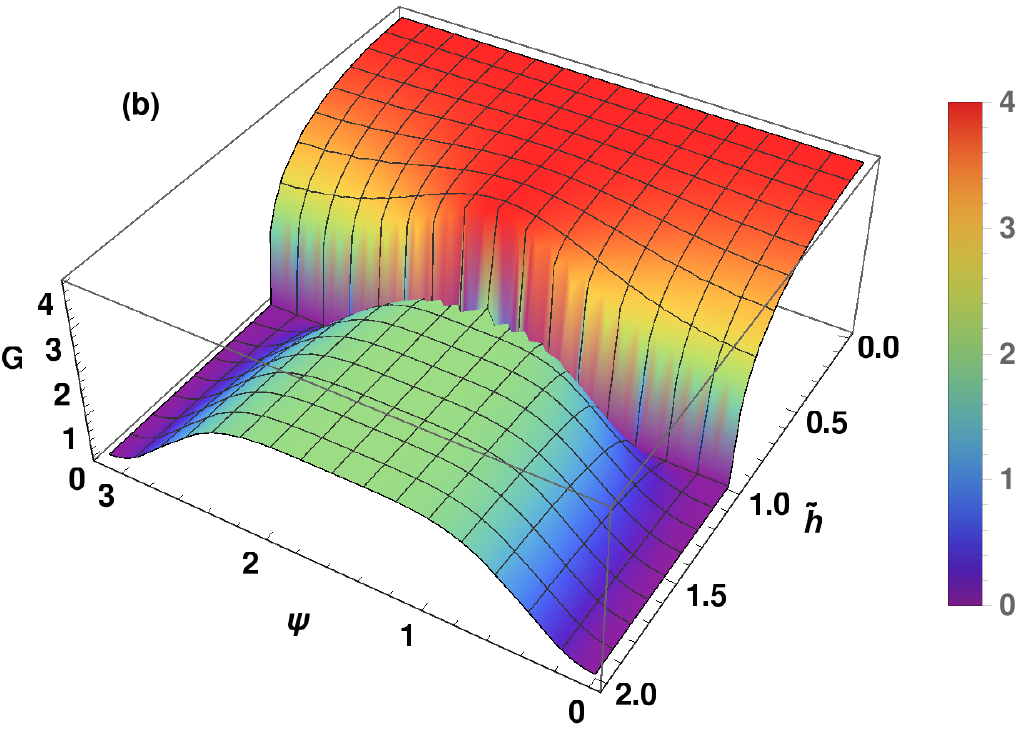}
\hspace*{\fill}%
\caption{(Color online) The features of differential conductance $G$ (in unit of $e^2/h$), at $\epsilon=0$, is demonstrated in $\tilde{h}$$-$$\Delta_p$ plane in panel (a) and $\tilde{h}$$-$$\psi$ plane 
in panel (b). We choose the values of the other parameters as $\psi=\pi/2$, $\mu=1$, $\Delta_s=1-\Delta_p$, $Z=4$, $U=15$ for panel (a) and $\mu=1$, $\Delta_s=0$, $\Delta_p=1$,
$Z=4$, $U=15$ for panel (b).
}
\label{G}
\end{figure*}
%---------------------------------------------------------------------------------------------------------------------
%---------------------------------------------------------------------------------------------------------------------

On the other hand, the wave-function inside the S region can be written as,
\bea
\Psi_S &=&c_1\,e^{ik_{S\ua}^{e}x}\{u_{\ua}(\theta_{+})e^{i\phi_S},0,0,\eta_{\ua}^{*}(\theta_{+})v_{\ua}(\theta_{+})\}^T\non \\
&&+c_2\,e^{ik_{S\da}^{e}x}\{0,u_{\da}(\theta_{+})e^{i\phi_S},\eta_{\da}^{*}(\theta_{+})v_{\da}(\theta_{+}),0\}^T\non\\
&&+d_1\,e^{-ik_{S\ua}^{h}x}\{\eta_{\ua}^{*}(\theta_{-})v_{\ua}(\theta_{-}),0,0,u_{\ua}(\theta_{-})e^{-i\phi_S}\}^T\non\\
&&+d_2\,e^{-ik_{S\da}^{h}x}\{0,\eta_{\da}^{*}(\theta_{-})v_{\da}(\theta_{-}),u_{\da}(\theta_{-})e^{-i\phi_S},0\}^T \ .
\eea
with $\eta_{\sigma}(\theta_\alpha)=s_\sigma\Delta_{\sigma}(\theta_\alpha)/
\lvert {\Delta_{\sigma}(\theta_\alpha)}\rvert$, $s_{\sigma}=(-1)^{\sigma-1}$ where $\sigma=\pm$\, denotes the $\ua,\da$ spin channels and 
$\alpha=\pm$ indicates the direction of motion. Momenta inside the F and S regions are given by: $k^{e/h}_{F\sigma}=\sqrt{2(\pm\epsilon+\mu+\sigma h)}$ and $k^{e/h}_{S\sigma}=\sqrt{2(\mu+U)\pm2\sqrt{\epsilon^2-|\Delta_\sigma|^2 }}$ respectively. Electron and hole components of the wave functions are given by,
$u_\sigma(\theta_\alpha)=
\frac{1}{\sqrt{2}}\Big(1+\frac{\sqrt{\epsilon^2- {\lvert\Delta_\sigma(\theta_\alpha)\rvert}^2}}{\epsilon}\Big)$ and 
$v_\sigma(\theta_\alpha)=
\frac{1}{\sqrt{2}}\Big(1-\frac{\sqrt{\epsilon^2- {\lvert\Delta_\sigma(\theta_\alpha)\rvert}^2}}{\epsilon}\Big)$ respectively.

Now using the proper boundary conditions, we find all the quantum mechanicl scattering amplitudes from the FS interface. At zero temperature, following the BTK formalism~\cite{blonder1982transition}, 
the differential conductance is given by, $G=G_0 \sum_{\sigma=\ua,\da}(1+R_{\sigma,\sigma}^A+R_{\sigma,-\sigma}^A- R_{\sigma,\sigma}-R_{\sigma,-\sigma} )$ where $G_0={2e^2 \over h} D(\theta) $ is the normal state conductance and $D(\theta)= 4 \cos^2 \theta /(Z^2+4 \cos^2 \theta)$~\cite{burset2014transport,paul2016transport}. Here, $R_{\sigma,\pm\sigma}(R_{\sigma,\pm\sigma}^A)$ are the reflection 
(AR) probability with conserved and flipped spin.

%----------------------------------------------------------------------
\section{ Results}{\label{sec:III}}
%----------------------------------------------------------------------
The behavior of the scattering probabilities are shown in Fig.~\ref{GvsE} as a function of the energy of the incident electron in the subgapped regime. Here, the F part of the NW is effectively in the half-metalic regime as $\tilde{h}\gg \mu$.
For simplicity, we have also chosen $\phi_F=0$ throughout our analysis.
For $\psi=0$, incoming electron spin is parallel to the $\bf{d}$ vector of the superconductor. This configuration hinders the possibility of cooper pairing and therefore only possible process is normal reflection
from the FS interface. Hence, the conductance vanishes in the subgapped regime.  For $\psi \neq 0$,  SFAR probability increases due to the dominance of $p$-wave pairing in the topological regime 
($\Delta_p>\Delta_s$). Note that, the general condition for gap closing is $\Delta_s=\Delta_p\cos{\theta}$ which reduces to $\Delta_s=\Delta_p$ in our case. On the other hand, for $\psi=\pi/4$, an anti-resonance in SFAR probability is clearly visible in Fig.~\ref{GvsE} at energy $\epsilon\sim0.3$ in the topological regime. This anti-resonance can be attributed to an interesting
interference between different spin channels when $\psi=\pi/4$.
Furthermore, in this parameter regime for any value of $\psi$, AR and spin flip reflections are prohibited due to the absence of other spin channel in the half metallic limit.

In Fig.~\ref{GvsEpsi}, we show the behavior of differential conductance G (in unit of $e^2/h$) in the plane of incident electron energy ($\epsilon$) and polarization angle $\psi$. We observe that G vanishes when incident electron spin is parallel (anti parallel) to the $\bf{d}$ vector \ie $\psi=0\, (\psi=\pi)$. This is consistent with 
the slice of Fig.~\ref{GvsEpsi}, denoted by the white dotted line and also depicted in Fig.~\ref{GvsE}. 
On the other hand, G reaches at its maximum value $2e^2/h$, at $\epsilon=0$, when the polarization of the ferromagnet is perpendicular to the $\bf{d}$ vector of the superconductor \ie $\psi=\pm\pi/2$. 
This occurs as SFAR due to triplet cooper pairing becomes maximum with this orientation. This is indicative of a single MZM contributing resonantly to transport.  
Note that, a pair of MZMs appear at the two ends of the superconductor in the topological phase and above observation implies that only a specific linear combination of pair of MZMs
is allowed to couple resonantly to the F region due to the spin selection rule while the other combination remains decoupled.
Such SFAR induced by MZM was earlier studied theoretically~\cite{he2014selective} and recently confirmed experimentally~\cite{sun2016majorana}. Furthermore, when the incident electron energy is comparable to the superconducting gap ($\epsilon\sim1$), reflection process dominates over AR. Hence, G becomes vanishingly small and independent of $\psi$ as can be seen from Fig.~\ref{GvsEpsi}. 
We also observe that the conductance peak splits as we move away from $\epsilon=0$, for a wide range of $\psi$ (see the highlighted black dashed line in Fig.~\ref{GvsEpsi}).
The ferromagnetic lead acts like a time-reversal breaking boundary perturbation to the pair of MZMs leading to their hybridization and hence resulting in split peaks.

The features of differential conductance $G$ (in unit of $e^2/h$), at $\epsilon=0$, is demonstrated in Fig.~\ref{G}(a) in $\Delta_p$$-$$\tilde{h}$ plane for $\psi=\pi/2$. Note that, when $\tilde{h}=0$ and
$\Delta_p>\Delta_s$ (topological regime), $G$ is $4e^2/h$ indicating two MZMs originating from the two different bands~\cite{burset2014transport} contributing resonantly to conductance. On the other hand, when $\tilde{h}>\mu$ \ie spin polarized regime and $\Delta_p>\Delta_s$, G is $2e^2/h$ indicating the fact that only one MZM participating in transport. Also, in the regime $\Delta_p>\Delta_s$, $G$ manifests a sharp transition from $4e^2/h$ to $2e^2/h$ with the variation of $\tilde{h}$ \ie depending on the availibility of the spin channels. At $\epsilon=0$, transport is being carried out solely by the subgapped MZMs and hence this sharp transition of $G$ indicates the spin selective coupling of MZM to the ferromagnetic lead which is the main result of our paper.
Furthermore, in the trivial regime ($\Delta_s>\Delta_p$), singlet cooper pairing is not possible owing to the blockage of one spin channel ($\tilde{h}>\mu$) and 
as a consequence $G$ vanishes. However, in the regime $\tilde{h}<\mu$, due to the availibility of both spin channels, normal AR gives rise to non-zero $G$. 

The sensitivity of zero energy conductance on polarization angle $\psi$, in the topological superconducting regime, can be seen from Fig.~\ref{G}(b). 
When the electron spin is perpendicular to the $\bf{d}$ vector \ie $\psi=\pi/2$, $G$ is $2e^2/h$ for the entire $\tilde{h}>\mu$ regime due spin selective coupling to one MZM. 
Around $\tilde{h}=\mu$, $G$ starts gradually increasing from $2e^2/h$ and finally reaches $4e^2/h$ for $\tilde{h}=0$ \ie both MZMs are resonantly coupled to the lead.
With the variation of $\psi$ from $\pi/2$ (towards $0$ or $\pi$), probability of SFAR decreases and hence $G$ decreases monotonically and becomes zero when $\psi=0$ and $\tilde{h}>\mu$. 
Hence, such spin dependent coupling of MZM, in the $\Delta_p>\Delta_s$ regime, explicitly depends on the polarization strength $\tilde{h}$ of the F regime and angle of magnetization $\psi$. 

%----------------------------------------------------------------------
\section{ Summary and Conclusions}{\label{sec:IV}}
%----------------------------------------------------------------------

To summarize, in this article, we study two terminal differential conductance of a quasi 1D FS junction where the superconductor consists of mixed singlet and triplet pairings. When the superconducting part of the NW becomes topological and $\tilde{h}$ is parallel (anti-parallel) to $\bf{d}$ vector ($\psi=0 (\psi=\pi)$), transport is blocked through the junction due to the absence of SFAR. On the other hand, when $\tilde{h}$ is perpendicular to $\bf{d}$ ($\psi=\pi/2$) differential conductance splits away from $\epsilon=0$ due to time-reversal breaking boundary perturbation. 
Moreover, zero energy conductance spectra shows sharp transition from $4e^2/h$ to $2e^2/h$ when $\tilde{h}>\mu$ \ie, as we move into the polarized regime.
Such transition between quantized conductances at zero bias demonstrates an efficient spin dependent coupling to a single MZM from the pair of MZMs, using a ferromagnetic lead.

In systems having proximity induced conventional superconductivity ($s$-wave), to realize MZMs at the two ends of a one-dimensional NW, the required ingredients 
are spin-orbit coupling in the NW and a magnetic field perpendicular to the spin-prbit field direction. The applied magnetic field and the chemical potential have to be tuned properly to 
achieve topological phase in the NW~\cite{mourik2012signatures,das2012zero}. On the other hand, in our set up, we do neither require spin-orbit coupling nor a zeeman gap to achieve 
topological phase hosting MZMs at each end of the superconducting part of the NW. Essentially, the relative magnitude of the intrinsic spin singlet and triplet pairings of the unconventional 
superconductor gives rise to the topological phase hosting pair of MZMs at the ends of the superconducting part of the NW. This motivates us to consider an inversion symmetry broken 
NCS type superconductor with mixed singlet and triplet pairings~\cite{yip2014noncentrosymmetric,smidman2017superconductivity} to study our model.  The strength of the stoner field $\bf{\tilde{h}}$ in the ferromagnetic probe can be thought off as an efficient way to controlling time reversal breaking boundary perturbation which leads to sharp transition of zero-bias differential conductance from the quantized value of $4e^2/h$ to $2e^2/h$ when the stoner field $\bf{\tilde{h}}$ in the ferromagnetic and the $\bf{d}$ vector of the superconductor are kept mutually perpendicular to each other. 

As far as practical realization of our model is concerned, a NW may be possible to fabricate in close proximity to a ferromagnet~\cite{hofstetter2010ferromagnetic} and NCS superconductor for \eg 
$\rm Mo_{3}Al_{2}C$, $\rm BiPd$ etc.~\cite{karki2010structure,mintumondal}. However, experimental realization of proximity induced unconventional superconductivity in 
semiconducting NW has not been reported so far, to the best of our knowledge. 
To validate our model, the orientation of the $\vec{d}$ vector of the spin-triplet component to be changed according 
to the direction of transport. Hence, the transport signatures (differential conductance $G$) must be measured in the plane orthogonal to the axis along which inversion symmetry broken 
spin-orbit coupling is present. Our proposed differential conductance spectra can be a probe for the spin selective coupling of pair of Majorana zero modes to the lead.

%-----------------------------------------------
{\it Acknowledgments.~}{GCP and AS acknowledge the warm hospitality of IISER Kolkata where part of the work was done.}
%-----------------------------------------------

%%%%%%%%%%%%%%%%%%%%%%%%%%%%%%%%%%%%%%%%%%%%%%%%%%%%%%

%-----------------------------------------------

%-----------------------------------------------
\bibliography{bibfile}{}
%-----------------------------------------------

\end{document}